\documentclass[12pt]{article}
\usepackage{epsfig}
\usepackage{amssymb,graphicx}
\def\be{\begin{equation}}
\def\ee{\end{equation}}
\def\ba{\begin{array}{c}}
\def\ea{\end{array}}

\def\ben{\[}
\def\een{\]}

\newcommand{\bea}{\begin{eqnarray}}
\newcommand{\eea}{\end{eqnarray}}

\begin{document}

%V_s(r)$ with $V_s(r)=

\begin{center}

{\Large \bf {

Symmetrized quartic polynomial oscillators and their partial exact
solvability
 }}

\vspace{13mm}

 {\bf Miloslav Znojil}

 \vspace{3mm}
Nuclear Physics Institute ASCR, Hlavn\'{\i} 130, 250 68 \v{R}e\v{z},
Czech Republic

 e-mail:
 %fruzicka@gmail.com
% %, ivezska@gmail.com
% and
  znojil@ujf.cas.cz

\vspace{3mm}

%\today, cubicl.tex

\end{center}

%\newpage
%, constructively,, constructively,

%\vspace{5mm}

\section*{Abstract}

Sextic polynomial oscillator is probably the best known quantum
system which is partially exactly {\it alias} quasi-exactly solvable
(QES), i.e., which possesses closed-form, elementary-function bound
states $\psi(x)$ at certain couplings and energies. In contrast, the
apparently simpler and phenomenologically more important quartic
polynomial oscillator is {\em not\,} QES. A resolution of the
paradox is proposed: The one-dimensional Schr\"{o}dinger equation is
shown QES after the analyticity-violating symmetrization
$V(x)=A|x|+B x^2+C|x|^3+x^4$ of the quartic polynomial potential.

\subsection*{Keywords:}

quantum bound states; non-numerical methods; piecewise analytic
potentials; quartic oscillators; quasi-exact states;

\subsection*{PACS:}

.

 03.65.Ge – Solutions of wave equations: bound states

% PACS 02.30.Gp – Special functions

02.30.Hq - Ordinary differential equations

\newpage
%and the formulation of the problem
% $L^2(\mathbb{R})$.

\section{Introduction\label{Introduction}}

A (certainly, not entirely complete) description of the dramatic
recent history of the discovery of existence of certain anomalous,
elementary, harmonic-oscillator-resembling exceptional non-numerical
bound-state solutions $\psi(x)$ of certain non-harmonic-oscillator
Schr\"{o}dinger equations may be found described in the dedicated
monograph by Ushveridze \cite{Ushveridze}. Although the author
himself calls these models quasi-exactly solvable (QES), he
immediately admits that the other authors may endow the same name
with a different meaning (this is a rather philosophical subtlety, a
deeper discussion of which may be found postponed to Appendix A). At
the same time, Ushveridze claims that irrespectively of the rigorous
definition of the QES concept, the popular quartic-oscillator
potentials
 \be
 V(x)=V^{(QO)}(x)=Ax+Bx^2+Cx^3+x^4\,.
 \label{quapo}
 \ee
{\em cannot} be considered QES.

The resulting absence of any non-numerical bound-state solutions of
one-dimensional Schr\"{o}dinger equations
 \be
 -\, \frac{{\rm d}^2}{{\rm d} x^2} \psi_n(x)
 + V(x) \psi_n(x)= E_n\,
 \psi_n(x)\,,\ \ \ \ \psi_n(x) \in L^2(\mathbb{R})\,
 %\ \ \ \ n = 0, 1, \ldots, N\ (N \leq \infty)
   \label{SEx}
  \ee
with interactions (\ref{quapo}) is to be perceived as strongly
unpleasant. Indeed, the simplified quantum model (\ref{quapo}) +
(\ref{SEx}) plays the phenomenologically extremely useful role in
quantum chemistry \cite{IJQC}. In parallel, it often serves as a
mathematical and methodical laboratory in relativistic quantum field
theory. {\it Pars pro toto}, let us just mention that at certain
couplings, Eq.~(\ref{quapo}) defines the radial part of the famous
Mexican-hat potential which samples the emergence of the
Nambu-Goldstone bosons in systems with spontaneously broken
symmetries (cf. picture Nr.~7 in \cite{Goldstone}).

Secondly, one must mention the context of atomic, molecular and
nuclear physics in which the double-well shape of the
one-dimensional version of the Mexican-hat potential (using, say, a
large negative $B\ll 0$ in (\ref{quapo})) proves useful as leading
to an apparent degeneracy of bound states with opposite parity.
Unfortunately, this is related to the quantum tunneling phenomenon
which is rather difficult to describe by non-numerical or even
semi-numerical means as sampled, e.g., by the conventional
perturbation theory (cf. also Refs.~\cite{DW,DWb,DWc} in this
context).

It seems worth adding that the Ushveridze's declaration of the
non-QES status of the quartic-oscillator potentials (\ref{quapo}) is
only valid under certain tacit assumptions including, e.g., the
conventional postulate that the potential should be analytic but, at
the same time, that the coordinate $x$ itself must be observable,
i.e., that it must lie strictly on the real line. After one abandons
one of the latter two assumptions, the Ushveridze's ``no-go''
theorem is not valid anymore. In the literature one can find several
counterexamples defined along a suitable complex curve of $x$ for
which the corresponding quartic oscillator becomes QES (cf., e.g.,
\cite{BBjpa,quarticZ}).

The discovery of the latter counterexamples was preceded by the
mathematical clarification of the consistent physical meaning of
having complex $x \in \mathbb{C}$ (cf. \cite{BG} or \cite{BT}). Such
an extension of the scope of the quantum model building also
inspired several phenomenologically oriented further developments
\cite{Mateo}. Nevertheless, we are not going to move along this line
here, mainly because the loss of the reality of the coordinate
changes the physics too much. In particular, the change implies such
a growth of extent of necessary additional mathematics \cite{book}
that the underlying QES philosophy of having a ``solvable'' model
gets, to a great extent, lost \cite{actasolva}.

This is the reason why we shall keep the coordinate $x$ real and
remove, instead, the other Ushveridze's tacit assumption of the
analyticity of the potential. In such a setting we shall reveal and
illustrate, constructively, the QES property of Schr\"{o}dinger
Eq.~(\ref{SEx}) with another quartic-polynomial-interaction
potential
  \be
 V(x)=A|x|+B x^2+C|x|^3+x^4\,
 \label{alquapone}
 \ee
which is made spatially symmetric ``by brute force'', i.e., at the
expense of the loss of its analyticity in the origin.

\section{Non-analytic QES models}

In the context of mathematical physics polynomials (\ref{quapo})
often play the role of benchmark interactions, say, in the tests of
perturbation expansions \cite{IJQC}. For this reason, as we already
indicated, it is not too fortunate that the related Schr\"{o}dinger
Eq.~(\ref{SEx}) must be solved, at {\em any} (real) triplet of
coupling constants $A$, $B$ and $C$,  by its brute-force numerical
integration.

\subsection{Inspiration: analytic sextic QES oscillators}

In the light of the above comment it sounds like a paradox that
elementary bound-state solutions do exist for the next,
higher-degree polynomial potentials
 \be
 V_{(sextic)}(x)=Ax^2+Bx^4+x^6\,,
 \label{sextpo}
 \ee
at some exceptional, {\it ad hoc} couplings and energies at least
\cite{Singh,Turbiner}. At the same time, in comparison with quartic
oscillators (\ref{quapo}), both the practical phenomenological
implementations and/or the methodical implications of the sextic
oscillators seem much less useful or relevant.

Let us add that the incomplete elementary solvability as exemplified
by the sextic polynomial interaction (\ref{sextpo}) is characterized
by the observation that a few exceptional bound states may still
possess an elementary, harmonic-oscillator-resembling product form
 \be
 \psi_n(x)=P(x)\,\exp W(x)
 \label{qesanza}
 \ee
where $P(x)=P(x,n)$ and $W(x)$ are suitable polynomials.

\subsection{Symmetrized WKB asymptotics}

In what follows we intend to favor, in a somewhat complementary
manner, the class of asymptotically fourth-power interactions
(\ref{alquapone}). Our guiding idea was that in contrast to the
partially solvable model (\ref{sextpo}) (and also in a way
reflecting the complicated Heun-function solvability of quartic
oscillators as mentioned in Appendix A below), the fully analytic
polynomial interaction potentials (\ref{quapo}) defined on the whole
real line are spatially asymmetric in general. This fact turned our
attention to the possible Hermitian modifications of the quartic
potentials and, in particular, to their spatially symmetrized
version (\ref{alquapone}).

The main specific merit of potentials (\ref{alquapone}) is that they
remain Hermitian and asymptotically analytic and confining, with the
analyticity merely violated in the origin. On this background one
only has to treat the point $x=0$ as a regular coordinate at which
the logarithmic derivatives of the bound-state wave functions
$\psi_n(x)$ must be matched in standard manner as known and tested
in the square-well models.

In what follows our main  task will lie in showing that and how the
entirely standard wave-function matching technique could be
implemented in the QES context. Our main conclusion will be that the
harmonic-oscillator-resembling elementary form of the quantum bound
states may fairly easily survive the loss of the analyticity of the
potential in the origin.

Polynomial $W(x)$ in the standard QES ansatz (\ref{qesanza}) remains
WKB-related and quantum-number-independent so that its form must be
in a one-to-one correspondence with the asymptotic behavior of the
potential. Naturally, there exists an immediate connection between
$W(x)$ and the WKB asymptotics of wave functions $\psi_n(x)$.
Nevertheless, only a polynomial form of $W(x)$ proves useful in the
QES context.

Its specification is routine. In particular, one may easily
construct polynomial $W$ for the above-mentioned analytic sextic
oscillators (\ref{sextpo}) yielding the proper
normalizability-guaranteeing polynomial $W_{(sextic)}(x) = -x^4/4 +
{\cal O}(x^3)$ (cf. \cite{Singh}). In contrast, one must be more
careful in the case of our present non-analytic quartic oscillators
(\ref{alquapone}). Branched, {\em non-analytic} asymptotics of any
candidate for a QES wave function are obtained,
 \be
 W_{(non-analytic)}(x)
 =
 \left \{
 \begin{array}{ll}
 +x^3/3+ {\cal O}(x^2)\,,&\ x \ll 0\,,\\
 -x^3/3+ {\cal O}(x^2)\,,&\ x \gg 0\,.
 \ea
 \right .
 \label{genera}
 \ee

\subsection{Wave-function matching in the origin}

In a search for elementary quartic-oscillator bound states we must
{\em necessarily} replace the analytic-function ansatz
(\ref{qesanza}) by its suitable weaker version. Once we start, say,
from the two-branched asymptotics (\ref{genera}) we may postulate,
e.g.,
 \be
 W_{(non-analytic)}(x)
 =
 \left \{
 \begin{array}{ll}
  W_{(left)}(x)=+x^3/3+ a\,x^2+b\,x\,,&\ x < 0\,,\\
 {W}_{(right)}(x)=-x^3/3+  \tilde{a}\,x^2-\tilde{b}\,x\,,&\ x >
 0\,.
 \ea
 \right .
 \label{kongenera}
 \ee
The related unavoidable loss of the analyticity of wave functions in
the origin implies that we will have to treat our model
(\ref{alquapone}) in a way similar to square wells, i.e., via
matching the logarithmic derivatives of $\psi_n(x)$ at the point of
non-analyticity, i.e., at $x=0$.

In the QES cases, in particular, we will have to assume that the
conventional ansatz (\ref{qesanza}) gets split into two branches,
 \be
 \psi_n^{(QES)}(x)
 =
 \left \{
 \begin{array}{ll}
 P_{(left)}(x,n)\,\exp W_{(left)}(x)\,,&\ x < 0\,,\\
 P_{(right)}(x,n)\,\exp W_{(right)}(x)\,,&\ x > 0\,,
 \ea
 \right .
 \label{rekongenera}
 \ee
where the form and construction of the two polynomials
$P_{(left/right)}(x,n)$ is still to be specified.

In general case, we may very easily replace, in addition, also the
analytic polynomial interactions (\ref{quapo}) by a broader class of
their generalizations which would be also manifestly non-analytic in
the origin and which could be rewritten in a six-parametric
two-branched form
 \be
 V_{(general)}(x)
 =
 \left \{
 \begin{array}{ll}
 V_{(general\ left)}(x)=Ax+Bx^2+Cx^3+x^4\,,&\ x < 0\,,\\
 {V}_{(general\ right)}(x)=\widetilde{A}x
 +\widetilde{B}x^2+\widetilde{C}x^3+x^4\,,&\ x > 0\,.
 \ea
 \right .
 \label{potgenera}
 \ee
This potential merely exhibits the left-right symmetry in asymptotic
domain so that we shall rather limit our attention to the spatially
symmetrized three-parametric potentials (\ref{alquapone}) re-written
in a slightly modified, piecewise-analytic notation,
 \be
 V^{(QES)}(x)
 =
 \left \{
 \begin{array}{ll}
 qx+rx^2+sx^3+x^4\,,&\ x < 0\,,\\
 -qx
 +rx^2-sx^3+x^4\,,&\ x > 0\,.
 \ea
 \right .
 \label{fipotgenera}
 \ee
 %
%the context of the search for the partial exact solutions
%
%\subsection{Symmetrized potentials}
%
%$V^{(QES)}(x)$
%\label{dvojkace} \label{secondary}}
%
The well known consequence of the spatial symmetry of our
Hamiltonians and potentials~(\ref{fipotgenera}) is that the wave
functions themselves must also be either symmetric or antisymmetric,
$\psi_n^{(QES)}(x)=\psi_n^{(even/odd)}(x)$. Thus, in the QES context
the key technical point is that it will be sufficient to stay, say,
just on the left half-axis with $x<0$. The resulting reduced QES
ansatz will then read
 \be
 \psi_n^{(even/odd)}(x)
 =\left(
 v_0^{(even/odd)}+v_1^{(even/odd)}x+\ldots+v_N^{(even/odd)}x^N
 \right )
 \exp W_{(left)}(x)\,,\ \ \ x<0\,.
 \label{anleft}
 \ee
Once we recall the necessary asymptotic boundary condition we get
 \be
 s=s(a)=4a=4\tilde{a}\,,
 \ \ \
 r=r(a,b)=4\,{a}^{2}+2\,b=4\,\tilde{a}^{2}+2\,\tilde{b}\,.
 \label{13}
  \ee
This means that we have to put $\tilde{a}=a$ and $\tilde{b}=b\,$ in
exponents (\ref{kongenera}).

The second requirement is the continuity of the wave function in the
origin. This means that we must put
$P_{(left)}(0,n)=P_{(right)}(0,n)$, i.e., $v_0^{(odd)}=0$ and,  say,
$v_0^{(even)}=1$ (= the choice of normalization). In parallel, the
continuity of the first derivative of the wave function in the
origin leads to the other constraint, viz., to the specification of
$v_1^{(even)}=-b$ or, say, $v_0^{(odd)}=1$ (= the choice of
normalization).

\section{QES constructions\label{dvojkabe}}

As long as the even-parity and odd-parity constructions remain
entirely analogous, the explicit constructive description of the
even solutions will be fully sufficient for our present illustrative
purposes. In a preparatory step let us restrict attention to a small
$N=2$.

\subsection{Even states $\psi_n^{(even)}(x)$ at $N=2$\label{ojkabe}}

%
%\subsection{QES constructions using {\em both} the wave functions
%and the potential matched in the origin}

Once we choose $N=2$ in our QES ansatz (\ref{anleft}) and denote
$P_{(left)}(x,n)=\left( 1+ux+v{x}^{2} \right)$, we may evaluate the
second derivative
%
%pro $z=1$ (sude) ted mame  SE: druha derivace ansatzu
 \be
 \psi''(x)={e^{1/3\,{x}^{3}+a{x}^{2}+bx}}\,\left [
 {x}^{6}v+ \left( 4\,av+u \right) {x}^{5}+ \left( 2\,bv+1+4\,{a}^{2}v+4
 \,au \right) {x}^{4}+ \right .
 \ee
 \ben
 +\left( 4\,abv+2\,bu+4\,a+4\,{a}^{2}u+6\,v
 \right) {x}^{3}+ \left( 10\,av+4\,u+4\,abu+{b}^{2}v+4\,{a}^{2}+2\,b
 \right) {x}^{2}+
 \een
 \ben
 \left .
 +\left( 4\,bv+2+{b}^{2}u+4\,ab+6\,au \right) x+2\,a+2
 \,bu+{b}^{2}+2\,v
 \right ]\,.
 \een
%zatimco
%prava strana je
In the light of Schr\"{o}dinger equation this must be equal to
expression $[V(x)-E]\,\psi(x)$ where we may put $E=-p$ and evaluate
 \be
 [V(x)+p]\,\psi(x)={e^{1/3\,{x}^{3}+a{x}^{2}+bx}}\,\left [
 {x}^{6}v+ \left( u+vs \right) {x}^{5}+
 \left( 1+us+vr \right) {x}^{4}+\right .
 \ee
 \ben
 +\left .
 \left( s+ur+vq \right) {x}^{3}+ \left( r+uq+vp \right) {x}^{2}+
 \left( q+up \right) x+p
 \right ]\,.
 \een
%takze srovnanim koeficientu u mocnin
The respective individual coefficients at powers ${x}^{5}$,
${x}^{4}$, \ldots, ${x}^{0}$ must equal each other. This comparison
generates the set of six algebraic equations
%v obou
%polynomech mame algebraicke rovnice
 \be
  -vs+4\,av  =0\,,\ \ \
   4\,{a}^{2}v+4\,au-vr-us+2\,bv
  =0\,,\ \ \
  %\right .
  \ee
  \ben
  %\left .
  4\,a-s+2\,bu+4\,{a}^{2}u+4\,abv-ur-vq+6\,v
 =0\,,\ \ \
 \een
 \ben
  -r+4\,abu+4\,u+{b}^{2}v+4\,{a}^{2}-vp+10\,av+
 2\,b-uq  =0\,,\ \ \
 \een
 \ben
  4\,bv-q+4\,ab+{b}^{2}u+2+6\,au-up
  =0\,,\ \ \ \ 2\,a+2\,bu+2\,v-p+{b}^{2}=0\,.
 \een
Due to relations (\ref{13}) the first two items are just identities
while the third one fixes the value of the last coupling constant,
%a posloupnost jejich reseni davajicich, as above,
% od 3 prvnich (tj. nejvyssich mocnin),
% vazbove konstanty potencialu,
 \be
 %s=s(a)=4a\,,
% \ \ \
% r=r(a,b)=4\,{a}^{2}+2\,b\,,
% \ \ \
 q=q(a,b)=4\,ab+6\,.
 \ee
%(notice that we have now selected $N=2$)
The value of the energy becomes determined by the fourth equation,
%ze ctvrte podminky
%pak dostaneme energii,
 \be
 E=E(a,b,v)=\frac {2\,u}{v}-10\,a-{b}^{2}\,.
 \ee
Once we construct just the even-parity state with property
%
%Pro symetricka reseni (licha vynechme) ted pridame pozadavek aby
$\psi'(0)=0$, i.e., $u=-b$,
%%. To pak znamena, ze spocitame
% \be
% \psi'(x)={e^{1/3\,{x}^{3}+a{x}^{2}+bx}} \left( {x}^{4}v+
% \left( u+2\,av \right) {x}^{3}+ \left( 1+2\,au+bv \right) {x}^{2}+
%  \left( 2\,a+2\,v+bu \right) x+u+b
% \right)
% \ee
%a musime pozadovat aby $u=-b$.
we are left with the last two algebraic equations. The first one
offers the two eligible wave-function-coefficient roots
%
%
%zbyvaji posledni dve  rovnice zjednodusime: Prvni ma dve reseni
 \be
v=v_\pm=
%1/4\,{\frac {2\,z+2\,au \pm 2\,\sqrt
%{{z}^{2}+2\,zau+{a}^{2}{u}^{2}-2\,b{u }^{2}}}{b}}
%%
%
%
 \frac{1}{4b}\,{ \left (-2\,ab+2 \pm 2\,\sqrt
{{a}^{2}{b}^{2}-2\,ab+1-2\,{b}^{3}}\right )}
 \ee
%a druha pak vede ke konecnemu zaveru, ze
while the last one yields
 \be
 a=a_\pm=\frac{1}{20\,b}
 \left( -{ {7}}\,{b}^{3}-8
 \pm { {3}}\,\sqrt {{b}^{6}-12\,{b}^{3}+16} \right)
 %
 %a=1/2\,{\frac {{b}^{5}}{-2+{b}^{3}}}_{{1}}
 \ee
and leaves just the real value of $b \neq 0$ independently variable.
%??????
%
%Zbyvajici volny parametr nam umoznuje chtit,
%aby potencial nemel spajk v pocatku (V'(0)=0).
%To dava podminku $2\,{ {{b}^{6}}/({-2+{b}^{3}}})+6$
%se dvema realnymi koreny
% \be
% b_\pm= (-3/2\pm 1/2\,\sqrt {33} )^{1/3}=
%                        1.111256679,
%                             -1.635194305
%                              \ee
%Pro prvni koren je uloha nezajimava (jama
%
%$V=-11.13359460+9.510603617\,x^2-5.399292642\,x^3+x^4$
%
%je jako HO, psi je jako gausian), pro druhy zajimava
%
%$V=13.52551431+.95531297e-1\,x^2+3.669288709\,x^3+x^4$
%
%(double well)
%
%je to druhy excitovany stav, ma nulu pro x=-.4106280215
%a pak hodne hluboky minimum
%
%%....

%\newpage
%********** Figure 2 zde
\begin{figure}                    %instead of \begin{figure}[t]
\begin{center}                         %instead of \begin{center}
\epsfig{file=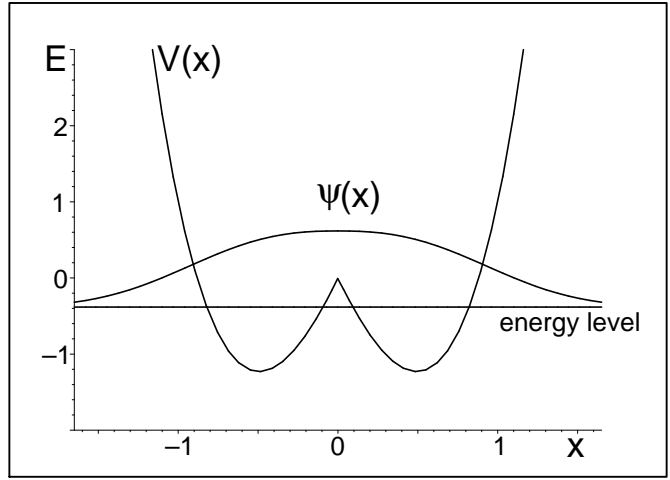,angle=270,width=0.50\textwidth}
\end{center}    % \sidecaption             %instead of \end{center}
\vspace{2mm} \caption{The QES ground state $\psi_0(x)$ in the $N=2$
quartic potential (\ref{fipotgenera}) at positive $b=1$. The effect
of spike in $V(x)$ proves negligible.
 \label{zvizt}
 }
\end{figure}

The resulting shapes of the QES potential (\ref{fipotgenera}) as
well as of the related exact wave function $\psi_n(x)$ are
illustrated in Figures \ref{zvizt} and \ref{nevizt} where we choose
$v=v_+$ and $a=a_+$. We picked up two sample values of $b=\pm 1$
and obtained the ground QES state and the second excited QES
state, i.e., solutions with $n=0$ and $n=2$, respectively.

%\newpage
%********** Figure 2 zde
\begin{figure}                    %instead of \begin{figure}[t]
\begin{center}                         %instead of \begin{center}
\epsfig{file=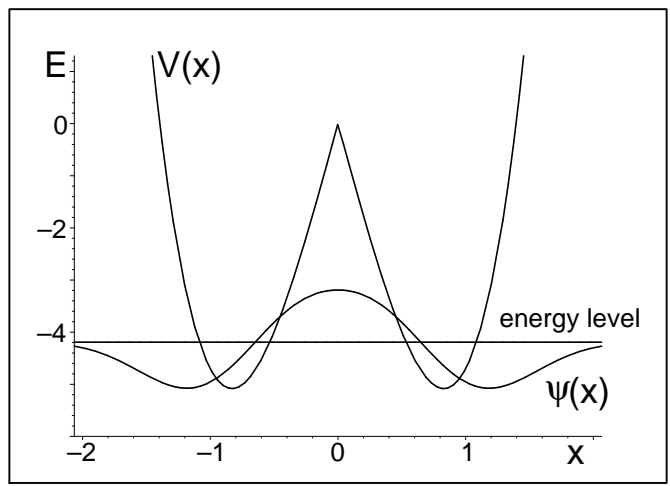,angle=270,width=0.50\textwidth}
\end{center}    % \sidecaption              %instead of \end{center}
\vspace{2mm} \caption{The second excited QES bound state $\psi_2(x)$
in the $N=2$ quartic potential (\ref{fipotgenera}) at negative
$b=-1$. The pronounced spike separates the potential, in effect,
into two almost decoupled wells.
 \label{nevizt}
 }
\end{figure}

We may conclude that although we sacrificed the analyticity of the
wave functions in the origin, the resulting potential is
defined by closed formula. At $b=1$ and $x<0$, for example,  we have
 \be
 V(x)=
   3\,\left( 1+1/\sqrt {5} \right) x+
    \left(2+ 9\, \left( -1+ {1}/\sqrt {5} \right) ^{2}/4
    \right) {x}^{2}
    + 3\,\left( -1+
 {1}/\sqrt {5} \right) {x}^{3}+{x}^{4}\,
  \label{exVesanza}
 \ee
i.e., after numerical
evaluation and symmetrization,
 \be
 V_{(QES)}(x)=
 \left \{
 \begin{array}{ll}
 4.3416\,x+2.6875\,x^2-1.6584\,x^3+x^4, & x<0\,,\\
 -4.3416\,x+2.6875\,x^2+1.6584\,x^3+x^4, & x>0\,.
 \ea
 \right .
 \label{Vesanza}
 \ee
We can summarize that the reconstruction of the $b-$parametrized
family of the $N=2$ QES potentials is feasible and friendly.
Numerically as well as non-numerically one can check the presence
and size of the spike in the origin or determine the position of the
local minima of the potential, etc. What remains to be added is the
discussion of the general case using any preselected integer $N$.

\subsection{General case \label{dvojkaa}}

Once we assume that $V(0)=0$ and that $x<0$ we may abbreviate
 \be
 V(x)-E=p+qx+r{x}^{2}+s{x}^{3}+{x}^{4}
 \label{trix}
 \ee
%- pri vhodnem zafixovani (libovolneho) pocatku shkaaly mame primo
%$E=-p$.
%
%
%asymptotics no problem
and differentiate our general polynomial Ansatz
 \be
 \psi(x)={e^{1/3\,{x}^{3}+a{x}^{2}+bx}}\, \sum_{k=0}^Nv_kx^k
 \ee
with $v_N \neq 0$ (and with the entirely formal definitions of
$v_{-1}=v_{N+1}=v_{N+2}=0$) yielding
 \be
 \psi'(x)=(x^2+2\,ax +b){e^{1/3\,{x}^{3}+a{x}^{2}
 +bx}}\, \sum_{k=0}^Nv_kx^k
 +{e^{1/3\,{x}^{3}+a{x}^{2}+bx}}\, \sum_{k=0}^N (k+1)\,v_{k+1}x^k
 \ee
and
 \be
 \psi''(x)=
 \left [(x^2+2\,ax +b)^2+2\,x+2\,a
 \right ]\,{e^{1/3\,{x}^{3}+a{x}^{2}+bx}}\, \sum_{k=0}^Nv_kx^k+
 \ee
 \ben
 +2\,(x^2+2\,ax +b){e^{1/3\,{x}^{3}
 +a{x}^{2}+bx}}\,\sum_{k=0}^N (k+1)\,v_{k+1}x^k
 +{e^{1/3\,{x}^{3}+a{x}^{2}+bx}}\,
 \sum_{k=0}^N (k+1)(k+2)\,v_{k+2}x^k\,.
 \een
The latter expression may be inserted in Schr\"{o}dinger equation
$\psi''(x)=(V(x)-E)\psi(x)$. In the resulting relation between
polynomials we already know that the contribution of the dominant
power $x^{N+3}$ fixes $s=s(a)=4a$, demonstrating the mutual
WKB-based large-coordinate correspondence between the subdominant
coupling and the subdominant exponent in the wave function.
Similarly we are also aware of the triviality of the coefficient at
the subdominant power $x^{N+2}$ yielding $r=r(a,b)=4\,{a}^{2}+2\,b$.
What is new is the consequence of the vanishing of the coefficient
at the sub-subdominant power $x^{N+1}$ which parametrizes also our
last QES coupling constant in an $N-$dependent manner,
$q=q(a,b,N)=4\,ab+2\,N+2$.

%\subsection{Algebraization}

%, i.e., the mutual WKB-based large-coordinate correspondence between
%the subdominant coupling and the subdominant exponent in the wave
%function

% (zbytek chceme symetrickej)

We are left with the set of $N+1$ recurrences for $N+1$ unknown
coefficients $v_k$. This set is most easily presented as the formal
eigenvalue problem
 \be
  \left (
 \begin{array}{cccccc}
 {\cal M}_{00}& {\cal M}_{01}& {\cal M}_{02}&  0&\ldots& 0\\
 {\cal M}_{10}& {\cal M}_{11}& {\cal M}_{12}&
 {\cal M}_{13}& \ddots&\vdots\\
 0& {\cal M}_{21}& {\cal M}_{22}& \ddots&  \ddots&0 \\
 0& \ddots& \ddots& \ddots&  {\cal M}_{N-2N-1}&  {\cal M}_{N-2N} \\
 \vdots&\ddots& 0& {\cal M}_{N-1N-2}&
 {\cal M}_{N-1N-1}& {\cal M}_{N-1N}\\
 0&\ldots&0& 0& {\cal M}_{NN-1}& {\cal M}_{NN}
 \ea
 \right )
  \left (
 \ba
 v_0\\
 v_1\\
 v_2\\
 \vdots\\
 v_N\ea
 \right ) = p\,\left (
 \ba
 v_0\\
 v_1\\
 v_2\\
 \vdots\\
 v_N\ea
 \right )\,
 \label{lefteig}
 \ee
where the matrix elements of the four-diagonal left-hand-side matrix
are just linear functions of parameters $a$ and $b$,
 \be
 {\cal M}_{k,k+2}=(k+1)(k+2)\,,
 \ \ \
 k=0,1,\ldots, N-2\,,
 \ee
 \be
 \ \ \
 {\cal M}_{m,m+1}=2\,b\,(m+1)\,,
 \ \ \
 {\cal M}_{m+1,m}=-2\,(N-m)\,,
 \ \ \
 m=0,1,\ldots, N-1\,
 \ee
and
 \be
 {\cal M}_{n,n}=4\,a\,n+2\,a+b^2\,,
 \ \ \
 n=0,1,\ldots, N\,.
 \ee
Whenever real, the $j-$th  eigenvalue may then represent an
auxiliary energy parameter,
 \be
 E=-p=-p_j(a,b,N)\,.
 \label{seculafr}
 \ee
Naturally, in the light of the illustrative $N=2$ example of
preceding paragraph the parameters $a$ and $b$ must guarantee the
wave-function matching in the origin. Thus, their values must be
determined as roots of the constraints, i.e., as roots of the
coupled pair of the above-discussed additional nonlinear equations
 \be
 v_0^{(even)}(a,b,N,j)=1\,, \ \ \ \ v_1^{(even)}(a,b,N,j)=-b
 \ee
(for the arbitrarily normalized even-parity QES states) or
 \be
 v_0^{(odd)}(a,b,N,j)=0\,,
 \ \ \ \ \ v_0^{(odd)}(a,b,N,j)=1\,
 \ee
(for the arbitrarily normalized odd-parity QES states).

\section{Summary\label{on}}

In summary, the use of branched asymptotics (\ref{kongenera}) +
(\ref{13}) and of the related non-analytic version
(\ref{rekongenera}) of ansatz (\ref{qesanza}) was shown here to lead
to the construction of an entirely new family of unconventional,
asymptotically quartic QES oscillators (\ref{alquapone}). They were
shown to share a number of useful closed-form-solvability properties
with their conventional, everywhere analytic sextic-oscillator
predecessors. In other words, we managed to prove that the
non-analytic but asymptotically quartic potential well
(\ref{alquapone}) may be perceived as quasi-exactly solvable.
Constructively we demonstrated that the potential offers a
bound-state model which is exactly, non-numerically solvable in
terms of polynomials in $x$ at certain couplings $A$, $B$, $C$ and
energies $E_n$.

The main technical ingredient which made the traditional QES recipe
perceivably (and also, for our purposes, {\em sufficiently}) more
flexible may be seen in our deliberate violation of the conventional
analyticity assumption at a single point, viz., at $x=0$. This
opened the possibility of making the potential (as well as the whole
Hamiltonian) spatially symmetric (i.e., ${\cal P}-$symmetric), with
the well known consequence of having also all of the bound states
$\psi_n(x)$ (and, in particular, also the exceptional QES states)
characterized by their even or odd parity, ${\cal
P}\,\psi_n(x)=\psi_n(-x)=\pm \psi_n(x)$.

The rest of our present story just forms a new, innovative but still
rather close parallel to the conventional (or, equally well, to the
above-mentioned less conventional, manifestly non-Hermitian)
linear-algebraic QES constructions as described in an extensive
dedicated literature (we may recommend the monograph
\cite{Ushveridze} as a source of further references).

Naturally,  in the future the implementation of the idea need not
remain restricted to the present, piecewise analytic quartic
interaction example. Still, we believe that due to an exceptional
methodical as well as practical (e.g., computation-testing) role of
the quartic-interaction class of models (with a single-point
non-analyticity in our present case) might lead to their quick
inclusion in the currently existing list of the available QES
quantum systems, with all of their valuable practical applications
as thoroughly reviewed, e.g., by Ushveridze \cite{Ushveridze}.

\subsection*{Acknowledgements}

Inspiring correspondence with Artur Ishkhanyan is gratefully
acknowledged. The work on the project was supported by the
Institutional Research Plan RVO61389005 and by the standard GA\v{C}R
Grant Nr. 16-22945S.

\newpage

\section*{Appendix A. A few comments on terminology}

%\subsection*{A.1. The grey zone between exact and numerical solvability}

In the literature the specification of the concept of the exact
solvability (ES) of Schr\"{o}dinger equations is often vague and
formulated {\it ad hoc}. For example, many people exclusively assign
the exceptional ES status to the simplest square-well models in
which the motion of a confined particle remains, locally, a free
motion. Another community of physicists admits solely analytic forms
of ES potentials $V_{}(x)$ requiring, in addition, that the ordinary
differential bound-state Schr\"{o}dinger Eq.~(\ref{SEx}) remains
solvable in terms of classical orthogonal polynomials \cite{Cooper}.

Whichever definition one accepts, the ES models remain separated
from the generic, purely numerical ones by an equally vaguely
specified grey zone in which, typically, one generalizes the
piecewise constant potentials $V(x)$ and takes advantage of a move
to the mere distributional, delta-function-like point interactions
\cite{Albeverio}. Alternatively, the separated, analytic-function
community perceives the grey solvability zone as covering, say, the
transition from the three-dimensional Coulomb potential
$V^{(CO)}(\vec{r}) = -e^2/|\vec{r}|$ (which is solvable in terms of
Laguerre polynomials) to its Ishkhanyan's \cite{Ishkh} long-range ES
modification $V^{(Ish)}(\vec{r}) = -e^2/\sqrt{|\vec{r}|}$ in
$s-$wave. The difference is that the latter model only proves
solvable in terms of  {\em non-terminating} confluent hypergeometric
functions. For this reason, the bound-state energies themselves
still do have just a numerical, non-ES, grey-zone (GZ) status.

%\subsection*{A.2. Quartic oscillators and the Heun's-function solvability}

Via an elementary change of variables the ES status of
$V^{(CO)}(\vec{r})$ (living on half-line) is shared with the
quadratic harmonic-oscillator polynomial interaction $V^{(HO)}(x)=
Ax+x^2$ living on the whole real line of Eq.~(\ref{SEx}). Similarly,
the {semi-numerical}, GZ solvability status of $V^{(Ish)}(\vec{r})$
is formally shared with the analytic quartic oscillator
(\ref{quapo}) on the line. This form of correspondence reflects the
reducibility of Eq.~(\ref{SEx}) + (\ref{quapo}) to the so called
Heun's differential equation \cite{Ishkh} which is just ``next'' to
the hypergeometric family and which still possesses a number of
exceptional GZ features \cite{Ronveaux}.

Inside such a GZ classification pattern the position of the
traditional QES class is fully inside the analytic-potential area.
For illustration people usually recall the formally privileged
status of the sextic model (\ref{sextpo}) while emphasizing that,
formally speaking, the model may be interpreted as an immediate
successor of harmonic oscillator as well as an immediate predecessor
of quartic oscillator of Eq.~(\ref{quapo}). The partial solvability
of the even-parity sextic oscillator makes it formally privileged in
comparison with the spatially asymmetric quartic polynomial
(\ref{quapo}) containing one more dynamics-determining coupling
constant.

Still, the conventional characterization of the three-parametric
quartic oscillators as ``purely numerical'' is not entirely
deserved, for several reasons. The main one has already been
mentioned above: The underlying ordinary differential
Schr\"{o}dinger equation belongs to the special class of Heun
equations \cite{Ronveaux}. Among the multiple benefits of using
these next-to-hypergeometric GZ equations we already mentioned the
recent discovery \cite{Ishkh} of the exact $s-$wave solvability of
bound states in the long-range central potential $V(\vec{r}) \sim
1/\sqrt{|\vec{r}|}$. Marginally, let us now add that the
phenomenological friendliness makes the quartic oscillators and
potential functions Eq.~(\ref{quapo}) rather popular even out of
quantum mechanics. Typically, they found applications even in the
theory of classical dynamical systems where the so called Lyapunov
function of the functional form (\ref{quapo}) can simulate one of
the most widespread bifurcation-evolution scenarios called ``cusp
catastrophe'' \cite{Zeeman,Zeemanb,Arnold}).

In 1998, an apparently impenetrable formal boundary between the
domains of quartic oscillators and of QES oscillators was broken by
Bender and Boettcher \cite{BBjpa}. They admitted the purely
numerical status of potentials (\ref{quapo}) but they discovered a
way out of the trap. In brief, they demonstrated, constructively,
that the elementary analytic solvability of the conventional (i.e.,
Hermitian) sextic model (\ref{sextpo}) {\em survives} the transition
to certain {modified}, non-conventional quartic-oscillator
Hamiltonians. The core of their proposal lied in the  replacement of
the real and confining potential (\ref{quapo}) by its complex plus
asymptotically ``wrong-sign'' alternative
  \be
 V(x)=V^{(BB)}(x)={\rm i}Ax+B x^2
 +{\rm i}Cx^3-x^4\,.
 \label{belpone}
 \ee
Although such a generalization already lies far beyond the scope of
our present paper (cf. also a few related comments in
\cite{BG,Mateo,book}), we only have to emphasize, in the conclusion,
that measured by the degree of solvability, analytic model
(\ref{sextpo}) and non-analytic model (\ref{belpone}) certainly
belong to the same (viz., QES) category, irrespectively of the
subtle details of its scope and rigorous definition.


\newpage

\begin{thebibliography}{00}

\bibitem{Ushveridze}
A. G. Ushveridze, Quasi-Exactly Solvable Models in Quantum Mechanics
(IOPP, Bristol, 1994).

\bibitem{IJQC}
Int. J. Quant. Chem. 21 (1982) (No. 1, special issue of Proceedings
of Int. Workshop on Pert. Theory at Large Order, edited by P.-O.
L\"{o}wdin and Y. \"{O}hrn), pp. 1 - 353.

\bibitem{Goldstone}
J. Goldstone,
 Il Nuovo Cimento %(1955-1965)
 19 (1961) 154.
%January 1961, Volume 19, Issue 1, pp 154-164
%Field theories with « Superconductor » solutions

\bibitem{DW}
%Double well model V (r) = ar2+br4+cr6 with a < 0 and perturbation
%method with triangular propagators
M. Znojil, Phys. Lett. A 222 (1996) 291.
%-8.

\bibitem{DWb}
%rD oscillators with arbitrary D > 0 and perturbation expansions with
%Sturmians.
M. Znojil, J. Math. Phys. 38 (1997) 5087.
%-97

\bibitem{DWc}
M. Znojil,
%Solvable simulation of a double-well problem in PT
%symmetric quantum mechanics (quant-ph/0303122),
J. Phys. A: Math. Gen. 36 (2003) 7639.
%-48.

\bibitem{BBjpa}
C. M. Bender and
 S. Boettcher,
 J. Phys. A: Math. Gen. 31 (1998) L273.
 %-L277

\bibitem{quarticZ}
M. Znojil,
%Quasi-exactly solvable quartic potentials with
%centrifugal and Coulombic terms (math-ph/0002036v2),
J. Phys. A: Math. Gen. 33 (2000) 4203.
% - 11.


\bibitem{BG}
%Mateo}
V. Buslaev and V. Grecchi,
% V. Equivalence of unstable anharmonic oscillators and double wells.
J. Phys. A: Math. Gen. 26 (1993) 5541.
%–5549.

\bibitem{BT}
%"Analytic Continuation of Eigenvalue Problems"
C. M. Bender and A. Turbiner, Phys. Lett. A 173 (1993) 442.

\bibitem{Mateo}
H. F. Jones and J. Mateo,
% J. Equivalent Hermitian Hamiltonian
%for the non-Hermitian -x4 potential.
Phys. Rev. D 73 (2006) 085002.

\bibitem{book}
F. Bagarello, J.-P. Gazeau, F. H. Szafraniec and M. Znojil (Ed.),
Non-Selfadjoint Operators in Quantum Physics: Mathematical Aspects
(John Wiley \& Sons, Hoboken, 2015).

\bibitem{actasolva}
M. Znojil,
%"New concept of solvability in Quantum Mechanics."
Acta Polytechnica 53 (2013) 473
%
%5): 473-482,
(http://dx.doi.org/10.14311/AP.2013.53.0473).
%(arXiv:1302.2035 [quant-ph])


\bibitem{Singh}
%Anharmonic oscillator and the analytic theory
%of continued fractions
%Virendra
V. Singh, S. N. Biswas and K. Datta,
Phys. Rev. D 18 (1978) 1901.

\bibitem{Turbiner}
 A. V. Turbiner, Comm. Math. Phys. 118 (1988) 467.

\bibitem{Cooper}
F. Cooper, A. Khare and U. Sukhatme,
% Supersymmetry and quantum mechanics.
Phys Rep 251 (1995) 267.
%–388.

\bibitem{Albeverio}
S. Albeverio, F. Gesztesy, R. Hoegh-Krohn and  H. Holden, Solvable
Models in Quantum Mechanics (Springer, New York, 1988).
%American Mathematical Soc.
%Book
%Texts and Monographs in Physics
%
%1988 Solvable Models in Quantum Mechanics
%Authors:Sergio Albeverio, Friedrich Gesztesy,
%Raphael Høegh-Krohn, Helge Holden
%… show all 4 hide
%ISBN: 978-3-642-88203-6 (Print) 978-3-642-88201-2 (Online)

\bibitem{Ishkh}
A. M. Ishkhanyan,
Exact solution of the Schrödinger equation for the inverse square
root potential $V_0/{\sqrt{x}}$.
Eur. Phys. Lett. 112 (2015) 10006,
arXiv:1509.00019.

\bibitem{Ronveaux}
A. Ronveaux (ed.), Heun's Differential Equations (Oxford Press,
London, 1995).

\bibitem{Zeeman}
%\bibitem{Thom}
R. Thom, Structural Stability and Morphogenesis. An Outline of a
General Theory of Models (Benjamin, Reading, 1975).

\bibitem{Zeemanb}
E. C. Zeeman, Catastrophe Theory - Selected Papers 1972–1977
(Addison-Wesley, Reading, 1977).

\bibitem{Arnold}
V. I. Arnold, Catastrophe Theory (Springer, Berlin, 1992).
%
%
%M. Znojil,
%%. Quantum catastrophes: a case study.
%J. Phys. A: Math. Theor. 45 (2012) 444036;

\end{thebibliography}
\end{document}